\documentclass[12pt]{article}
\usepackage{graphicx}
\newcommand{\preprint}[1]{\begin{flushright}#1\end{flushright}}

\newcommand \bra[1]{\left< {#1} \,\right\vert}
\newcommand \ket[1]{\left\vert\, {#1} \, \right>}

\newcommand{\bea}{\begin{eqnarray}}
\newcommand{\eea}{\end{eqnarray}}
\newcommand{\simgt}{\hbox{ \raise3pt\hbox to 0pt{$>$}\raise-3pt\hbox{$\sim$} }}
\newcommand{\simlt}{\hbox{ \raise3pt\hbox to 0pt{$<$}\raise-3pt\hbox{$\sim$} }}

\newcommand{\clfn}{\setcounter{footnote}{0}}

\def\als{\alpha_{\rm s}}
\def\lQ{\Lambda_{\rm QCD}}

\begin{document}
\preprint{TU--607\\ IFUM--677--FT\\ January 2001}
\vspace*{3cm}
\begin{center}
  {\bf\Large
Quarkonium Spectroscopy and Perturbative QCD
:\\
\vspace{4mm}
A New Perspective
}
  \\[20mm]
  {\large
    N.~Brambilla$^1$, Y.~Sumino$^2$ and A.~Vairo$^3$
    }
  \\[10mm]
  {\it
    $^1$
    Dipartimento di Fisica dell'Universit\`a di Milano  \\
    via Celoria 16, 20133 Milano, Italy
    }
  \\[10mm]
  {\it
    $^2$
    Department of Physics, Tohoku University\\
    Sendai, 980-8578 Japan
    }
  \\[10mm]
  {\it
    $^3$
    Institut f\"ur Theoretische Physik, Universit\"at Heidelberg\\
    Philosophenweg 16, D-69120 Heidelberg, Germany
    }

\end{center}
\vspace{1cm}
\begin{abstract}
\small
We study the energy spectrum of bottomonium in perturbative QCD, 
taking $\als(M_Z)=0.1181 \pm 0.0020$ as input and fixing 
$m_b^{\overline{\rm MS}}(m_b^{\overline{\rm MS}})$ on the  $\Upsilon(1S)$ mass.
Contrary to wide beliefs, perturbative QCD reproduces reasonably well 
the gross structure of the spectrum as long as the coupling constant remains smaller than one. 
We perform a detailed analysis and  discuss the size of non-perturbative effects.
A new qualitative picture on the structure of the bottomonium spectrum is provided. 
The lowest-lying $c\bar{c}$ and $b\bar{c}$ states are also examined.
\end{abstract}

\newpage

\section{Introduction}
\label{s1}
In recent years our knowledge of heavy quarkonia in the framework of perturbative QCD 
has developed significantly. On one hand computations of new higher-order corrections 
to various physical quantities appeared \cite{n3lo,mce,ks,hoangcm}. On the other hand the discovery 
of the mechanism of the renormalon cancellation in the quarkonium spectrum 
\cite{renormalon1,renormalon2} led to a drastic improvement of the convergence of the perturbative 
expansion of the energy levels.  As important applications up to date, the theory enabled precise 
determinations of the $\overline{\rm MS}$-mass of the bottom quark \cite{py,pp,my,upsilonmass}
and (in the future) of the top quark \cite{topcollab} from (mainly) the energy spectra 
of the lowest-lying states. The main uncertainty comes, in the bottomonium case, 
from the (essentially) unknown non-perturbative contributions. These are generally claimed 
to be around 100 MeV and ultimately set the precision of the prediction. 

As for the structure of the higher quarkonium levels, the Coulomb potential seems to be unable 
to explain it (for a review see \cite{hugs}). For instance, the spacing between two consecutive 
$nS$ levels does not decrease with $n$ and appears roughly  constant. For this reason several 
confining potentials have been introduced in the literature over the last decades. 
We refer to \cite{eq} for one of the most recent phenomenological study on several of them. 
Progress in this direction seems to be no longer possible, as far as the relation of the confining 
potentials with QCD remains obscure. This relation has been elucidated recently in \cite{long,m12} where 
a complete and systematic parametrization inside QCD of perturbative and non-perturbative effects 
of the heavy quarkonium interaction has been realized.
The parametrization depends on the mutual relation between the scale of non-perturbative physics, $\lQ$, 
and the other dynamical scales in the specific heavy quarkonium system under study.
A way to determine it and the size and nature of the non-perturbative contributions, 
consists in establishing to which extent perturbative QCD can consistently and successfully 
describe the system. This is the aim of the present paper, which investigates the range of validity 
of perturbative QCD on the heavy quarkonium spectroscopy and consequently extracts upper bounds 
to the non-perturbative contributions by comparing the perturbative predictions, 
at the current best accuracy, with the experimental data.

One of the main problems in having a consistent (i.e. convergent) perturbative 
expansion for the quarkonium energy levels has been for a long time its bad behaviour 
when expressed in terms of the pole mass (see, for instance, 
the poor convergence of the computations reported in \cite{yn3,py}). 
The reason has been understood in the renormalon 
language: the pole mass \cite{bbb} and the QCD static potential \cite{al}, respectively, 
contain renormalon contributions of order $\lQ$, which get cancelled in the total energy of a color singlet
quark-antiquark system \cite{renormalon1,renormalon2}.
The solution then consists in making explicit this cancellation by substituting in the energy expansion  
a short-range mass (free from infrared ambiguities) for the pole mass.
In \cite{upsilonmass,my,hoangcm} this approach has been used in order to calculate the bottom quark 
mass by means of sum rules. However, an analysis of the consistency 
between the whole level structure of perturbative QCD and the experimental data is lacking and 
will be done in the present work. In order to make explicit the renormalon cancellation 
we will express the quarkonium energies in terms of the $\overline{\hbox{MS}}$ masses 
and rearrange the perturbative series in the so-called $\varepsilon$ expansion \cite{hlm}. 
This will be the key ingredient of our analysis. We note that according to a formal argument, 
the residual uncertainty of the perturbative expansion due to the next-to-leading renormalon 
contribution is estimated to be  $\Lambda_{\rm QCD}\times (a_X \Lambda_{\rm QCD})^2$ for a bound-state 
$X$ of size $a_X$ \cite{al,long}, e.g.\ it amounts  to 5--20~MeV for the $1S$ bottomonium state. 
Provided this argument applies, in principle the predictions of perturbative QCD can be made 
precise down to this accuracy by sufficiently increasing the orders of the perturbative expansion.

The paper is organized as follows. In Sec. \ref{s2} we set up the formalism and the computational 
strategy. In Sec. \ref{s3} we perform the numerical analysis and in Sec. \ref{s4} we discuss the 
errors. In Sec. \ref{s5} we interpret our result and in Sec. \ref{s6} we draw some conclusions.

\section{Perturbative Expansions}
\label{s2}

We assume that the system is non-relativistic, $v\ll 1$, where
$v$ is the typical size of the heavy quark velocity in a quarkonium state.
Also, we assume that the soft scale $m v$ is much larger than $\lQ$.
Under these assumptions,
the energy (mass) of a quarkonium state $X$ is given as a series expansion
in the $\overline{\rm MS}$ coupling constant $\als(\mu)$
defined in the theory with $n_l$ massless quarks only.
We write a general expression valid also when the masses of the quark and 
antiquark are different. 
The reduced mass is defined from the pole masses by
$m_r = m_{1,{\rm pole}}m_{2,{\rm pole}}/(m_{1,{\rm pole}}+m_{2,{\rm pole}})$. 
Moreover, we define $x = 1-4m_r/(m_{1,{\rm pole}}+m_{2,{\rm pole}})$;
when the two masses are equal, $x = 0$. Up to ${\cal O}(\als^4 m)$
the energy of a heavy quarkonium state $X$, identified by the  quantum numbers 
$n,l,s$ and $j$, is given by\footnote{
The full formula up to ${\cal O}(\als^4 m)$
for the $S$ state spectrum was derived in \cite{py} and later
confirmed in \cite{my};
additional corrections necessary for the spectrum of $l \geq 1$ states
can be found in \cite{yn3};
the formula for the unequal mass case was derived in \cite{bcbv}.
}
\bea
&&
E_{X}(\mu,\als(\mu),m_{i,{\rm pole}}) =  m_{1,{\rm pole}}+m_{2,{\rm pole}} 
+ E_{{\rm bin},X}(\mu,\als(\mu),m_{i,{\rm pole}}) ,
\label{spectrum}
\\ &&
E_{{\rm bin},X}(\mu,\als(\mu),m_{i,{\rm pole}}) = - \, \frac{8}{9n^2} \, \als(\mu)^2  m_{r}
\sum_{k=0}^2 \varepsilon^{k+1} \left( \frac{\als(\mu)}{\pi} \right)^k P_k (L_{nl}) ,
\label{spectrum2}
\eea
where $\varepsilon=1$ is the parameter that will be used in order to properly organize 
the perturbative expansion in view of the ${\cal O}(\lQ)$ renormalon cancellation. 
$P_k(L_{nl})$ is a $k$-th-degree polynomial of $\displaystyle L_{nl} 
\equiv \log[\, 3n\mu / (8 \als(\mu)m_{r}) \, ] +S_1(n+l)+\frac{5}{6}$, and
the harmonic sums are defined as $\displaystyle S_p (q) \equiv \sum_{k=1}^{q} \frac{1}{k^p}$.
It is convenient to decompose the polynomials into renormalization-group
invariant subsets:
\bea
P_0 &=& 1,
\\
P_1 &=& \beta_0  \, L_{nl} + c_1 ,
\\
P_2 &=&
\frac{3}{4} \beta_0^2 \, {L_{nl}^2} + 
  \left( - \frac{1}{2} \beta_0^2 
             + 
     {\frac{1}{4}\beta_1 } + 
     {\frac{3}{2} \beta_0 {c_1}} \right) L_{nl}  + 
  {c_2} .
\eea
Here, $\beta_k$'s denote the coefficients of the QCD beta function, 
$\beta_0 = 11 - 2 n_l/3 $,
$\beta_1 = 102 - 38 n_l/3$, and $c_k$'s are given by
\bea
c_1&=& -4 ,
\\
c_2 &=&
- 
  {\frac{16\,{{\pi }^2}\,\left\{ 2s ( s+1 ) 
         ( 1 - x )  + 3 x \right\} }{27\,n}} 
\,     \delta_{l0}
- 
  {\frac{8\,{{\pi }^2}\,\left( D_S + 3\, X_{LS} \right)  }{9\,n\,l\,
      \left( l+1 \right) \,\left( 2\,l +1 \right) }} 
\,
      \left( 1 \! - \! \delta_{l0} \right)
+ 
\beta_0^2 \,\nu (n,l) 
\nonumber \\&&
- 
  {\frac{\left( 11 + x \right) \,{{\pi }^2} }{9\,{n^2}}} 
+ 
  {\frac{68\,{{\pi }^2}}{9\,n\,\left( 2\,l +1\right) }}
+
{\frac{473}{16}} + {\frac{9\,{{\pi }^2}}{2}} 
+ {\frac{33\,\zeta_3}{4}} 
- 
  {\frac{9\,{{\pi }^4}}{32}} 
- 
  {n_l}\,\left( {\frac{109}{72}} +
     {\frac{13\,\zeta_3}{6}} \right) ,
\nonumber \\ &&
\eea
with
\bea
&&
D_{S} \equiv
\left< 
3 \frac{(\vec{r}\cdot \vec{S})^2}{r^2} - \vec{S}^2 
\right>
=
\frac{
2 l (l+1) s (s+1) - 3 X_{LS} - 6 X_{LS}^2
}{
(2l-1)(2l+3)
},
\\&&
X_{LS} \equiv
\left< \vec{L}\cdot \vec{S} \right>
= \frac{1}{2}\,
\left[ j(j+1)-l(l+1)-s(s+1) \right] ,
\\&&
\nu (n,l)=
\frac{\pi^2}{8}
- \frac{1}{2} \, S_2(n+l)
+ \frac{n}{2} \frac{(n+l)!}{(n-l-1)!}
\sum_{k=1}^{\infty}
\frac{(n-l+k-1)!}{(n+l+k)! \, k^3}
\nonumber \\&& 
~~~~~~~~~~~~
+ \frac{(n-l-1)!}{2(n+l)!}
\sum_{k=1}^{n-l-1}
\frac{(2l+k)! \, (2k+2l-n)}
{(k-1)! \, (k+l-n)^3} .
\label{nu}
\eea
It is understood that the last term of Eq. (\ref{nu}) is zero
if $n-l<2$.\footnote{The infinite sum in Eq.~(\ref{nu}) can be easily evaluated 
analytically in terms of $\zeta_3$, etc.  for given values of $n$ and $l$, e.g. by using {\it Mathematica}.}

Next we rewrite the series expansion of $E_{X}$ in terms of the $\overline{\rm MS}$ masses. 
This is done by expressing the pole masses $m_{i,{\rm pole}}$ in terms of the  
renormalization--group-invariant $\overline{\rm MS}$ masses 
$\displaystyle \overline{m}_i \equiv m_{i,\overline{\rm MS}} (m_{i,\overline{\rm MS}})$:
\begin{eqnarray}
m_{i,{\rm pole}} = \overline{m}_i \left\{ 1 + {4\over 3} \varepsilon {\als(\overline{m}_i)\over \pi} 
+ \varepsilon^2  \left({\als(\overline{m}_i)\over \pi}\right)^2 d_1 
+ \varepsilon^3 \left({\als(\overline{m}_i)\over \pi}\right)^3 d_2 \right\}.
\label{pole}
\end{eqnarray}
The coefficient $d_1$ is given in \cite{gbgs}, while the analytic expression of $d_2$ can be
derived from the result of \cite{polemass2}, the renormalization-group evolutions of $\als(\mu)$
and $m_{\overline{\rm MS}}(\mu)$, and the matching condition \cite{lrv}.
Note that the counting in $\varepsilon$ in Eq.~(\ref{spectrum2}) and Eq.~(\ref{pole}) 
does not reflect the order in $\als$ but the wanted renormalon cancellation.
One way to understand this is to consider that in the sum of the pole-quark masses and the 
static QCD potential, $\sum_i m_{i,{\rm pole}} + V_{\rm QCD}(r)$, the renormalon cancellation takes 
place without reordering of power counting in $\als$ \cite{renormalon1,renormalon2}. 
The extra power of $\als$ comes in the energy level expansion when the dynamical variable $r^{-1}$ 
is replaced by the dynamical scale $\bra{nlsj} r^{-1} \ket{nlsj} \sim C_F \als m_r / n$.\footnote{
Since the soft scale is, in perturbative QCD,  the only scale involved
in the calculation of the binding energy, $E_{{\rm bin},X}$, up to ${\cal O}(\als^4 m)$, 
we may absorb the extra $\als$ by reexpressing $E_{{\rm bin},X}$ in terms of $\als(\mu)$ and
the soft scales $\mu_{i,{\rm soft}} = \als(\mu) \, \overline{m}_i$. Then the order counting in $\als(\mu)$ 
coincides with the order counting in $\varepsilon$ given here.}
Moreover, in order to realize the renormalon cancellation at each order of the
expansion, it is necessary to expand $m_{i,{\rm pole}}$ and $E_{{\rm bin},X}$ 
in the same coupling \cite{mar,hlm,sumino}, therefore 
we express $\als(\overline{m}_i)$ in (\ref{pole}) in terms of $\als(\mu)$: 
\begin{equation}
\als(\overline{m}_i)= \als(\mu) \left\{ 
1 + \varepsilon \, \frac{\als(\mu)}{\pi}  \,\frac{\beta_0}{2} \log\left(\frac{\mu}{\overline{m}_i}\right) 
+ \varepsilon^2 \left( \frac{\als(\mu)}{\pi} \right)^2 \,
\Biggl[ \frac{\beta_0^2}{4} \, \log^2 \left(\frac{\mu}{\overline{m}_i}\right)
+ \frac{\beta_1}{8} \log\left(\frac{\mu}{\overline{m}_i}\right) \Biggr] \right\}
.
\label{alphams}
\end{equation}
Substituting Eqs. (\ref{alphams}) and (\ref{pole}) into Eqs. (\ref{spectrum2}) and (\ref{spectrum}) 
we get an expression for the energy levels of the heavy quarkonium states, 
which depends on $\mu$, $\als(\mu)$ and $\overline{m}_i$, that we can organize as an expansion
in $\varepsilon$ up to order $\varepsilon^3$:
\begin{eqnarray}
E_{X}(\mu, \als(\mu),\overline{m}_i) &=& \overline{m}_1 + \overline{m}_2 
+ E_{X}^{(1)}(\mu, \als(\mu),\overline{m}_i) \varepsilon 
+ E_{X}^{(2)}(\mu, \als(\mu),\overline{m}_i) \varepsilon^2 \nonumber\\
& & + E_{X}^{(3)}(\mu, \als(\mu),\overline{m}_i) \varepsilon^3 + \dots
\label{en2}
\end{eqnarray}
Since the counting in $\varepsilon$ explicitly realizes the order $\lQ$ renormalon cancellation, 
and since $\als$ and $\overline{m}_i$ are short-range quantities, the obtained perturbative expansion 
(\ref{en2}) is expected to show a better convergence with respect to the original 
expansion (\ref{spectrum2}).

The obtained quarkonium mass $E_X$ depends on the scale $\mu$, due to our incomplete knowledge 
of the perturbative series. We fix the scale $\mu$ by demanding stability against variation of the scale:
\bea
\left.\frac{d}{d\mu} E_{X}(\mu, \als(\mu),\overline{m}_i)\right|_{\mu = \mu_X} = 0 .
\label{scalefix}
\eea
When we do this, we expect that the convergence properties of the series become optimal, and
that the scale becomes close to the inverse of the physical size of the bound-state $X$.
If the scale fixed by Eq. (\ref{scalefix}) evidently does not fulfill these expectations,  
then the theoretical predictions obtained in this way will be considered {\it unreliable}.
We will show that this typically happens when the coupling constant becomes bigger than one.

Based on the formalism developed in \cite{long},
in principle three scenarios are possible under the assumption
$v\ll 1$ and $mv \gg \lQ$:~
1) The energy or ultrasoft scale $m v^2$  is much larger than  $\lQ$. 
In this case the potential is entirely perturbative and non-perturbative corrections 
are parameterized by local condensates of the Voloshin--Leutwyler type \cite{vol}.  
2) The energy scale $m v^2$  is of order  $\lQ$. 
The potential is entirely perturbative and non-perturbative corrections are parameterized by non-local condensates.
3) The energy scale, $m v^2$, is smaller than $\lQ$. 
In this case short-range non-perturbative corrections affect the potential. They are parameterized 
by non-local condensates. 
A perturbative treatment of the energy levels is consistent
only if these are small compared to the Coulomb potential.
Explicit formulas for the non-perturbative contributions in all these three 
cases can be found in \cite{long}. 
However, their sizes in an actual calculation of the heavy quarkonium spectrum
are affected by large uncertainties. 
The best known case is 1): 
The leading non-perturbative contribution is 
$\simeq 1.44 \, n^6 \langle \als F^2(0) \rangle / (m^3 \als(\mu)^4)$.
It runs out of control for $n>1$, so the estimate is suitable only for 
the heavy quarkonium ground states. 
It is quite sensitive to the value of the gluon condensate and of $\als$. 
It is also numerically sensitive to the replacement of the strong coupling 
constant 
in the $\overline{\rm MS}$ scheme in the denominator with the coupling constant 
in some other scheme (e.g. the $V$ scheme or similar) as used by some authors \cite{py,yn3}.
Non-local condensates, which are suitable for the situations 2) and 3), 
are even less known. 
Moreover, they get entangled with higher-order 
perturbative corrections (starting at order $\als^5 m \log\als$ \cite{n3lo})
and in particular with the ${\cal O}(\lQ^3)$ renormalon \cite{long}.
Due to these uncertainties a direct evaluation of non-perturbative and higher-order contributions 
will not be included in the present investigation. 
Neither we will distinguish among the three different scenarios outlined above. 
Our approach consists in looking only at the perturbative spectrum up to 
${\cal O}(\als^4 m)$, 
under the general assumption that $mv \gg \lQ$.  
The internal consistency of the perturbative series (i.e. its convergence) 
and the comparison with the experimental data will provide indirect, but more stringent, contraints on 
the size of non-perturbative and higher-order corrections.

\section{Numerical Analyses}
\label{s3}
In this section we examine the series expansions of the quarkonium spectra, Eq.~(\ref{en2}), numerically.
The input value is $\als^{(5)}(M_Z) = 0.1181$ \cite{pdg}. We evolve the coupling and match 
it to the couplings of the theory with $n_l=4$ and 3 successively following \cite{running}\footnote{
We take the matching scales as $\overline{m}_b$ and $\overline{m}_c$, respectively.} 
and solving the renormalization-group equation numerically (4-loop running). 
In this section we do not take into account effects of the non-zero charm mass 
in the $b\bar{b}$ and $b\bar{c}$ systems. 

Since we expect the ground states of the $b\bar{b}$ and $c\bar{c}$
systems to be the states less affected by non-perturbative corrections,
we fix $\overline{m}_b$ and $\overline{m}_c$ through the conditions
\begin{eqnarray}
E_{\Upsilon(1S)} (\mu_X, \als(\mu_X),\overline{m}_b) &=& E_{\Upsilon(1S)}^{exp} = 9.460 \, \hbox{GeV} 
\label{con1},\\
E_{J/\psi} (\mu_X, \als(\mu_X),\overline{m}_c) &=& E_{J/\psi}^{exp} = 3.097 \, \hbox{GeV} , 
\label{con2}
\end{eqnarray}
where the experimental values of the vector ground states have been taken from \cite{pdg}.  
We assume, for the moment, that this identification is not affected by non-perturbative corrections. 
From Eqs. (\ref{con1}), (\ref{con2}) and (\ref{scalefix}) we determine $\mu_X$ 
(see Tab.~\ref{table:spectra}), and the $b$ and $c$ $\overline{\rm MS}$ masses:
\bea
\overline{m}_b \equiv
{m}_b^{\overline{\rm MS}}({m}_b^{\overline{\rm MS}})=4.203~{\rm GeV} ,
\label{bmass}
\\
\overline{m}_c \equiv
{m}_c^{\overline{\rm MS}}({m}_c^{\overline{\rm MS}})=1.243~{\rm GeV} .
\eea
These values are in good agreement with recent estimates based on $\Upsilon$ \cite{my,upsilonmass} and 
charmonium \cite{jamin} sum rules respectively.  

\begin{table}
\begin{center}
\begin{tabular}{c|r|r|c|l|r|l|l|l}
\hline
State $X$ &$E_X^{exp}~~$ &$E_X~~~$ & $E_X^{exp}-E_X$ &
$E_X^{(1)}$ &$E_X^{(2)}$ &$E_X^{(3)}$ &~~$\mu_X$ 
&$\als(\mu_X)$\\ 
\hline
$J/\psi$       & 3.097~~ & 3.097~  & 0 & 0.362 & 0.205 & 0.043  & 1.07 & 0.448\\
$\eta_c (1^1S_0)$ & 2.980~~ & 3.056~ & $-0.076$ & 0.333 & 0.195 & 0.042 &
1.23 & 0.399 \\
\hline
$\Upsilon(1^3S_1)$  &  9.460~~ & 9.460~ & 0        & 0.837 & 0.204  & 0.013 & 2.49 & 0.274  \\
$\Upsilon(1 ^3P_0)$ &  9.860~~ & 9.905~ & $-0.045$ & 1.38  & 0.115  & 0.003 & 1.18 & 0.409  \\
$\Upsilon(1 ^3P_1)$ &  9.893~~ & 9.904~ & $-0.011$ & 1.40  & 0.098  & 0.002 & 1.15 & 0.416  \\
$\Upsilon(1 ^3P_2)$ &  9.913~~ & 9.916~ & $-0.003$ & 1.42  & 0.086  & 0.003 & 1.13 & 0.422  \\
$\Upsilon(2 ^3S_1)$ & 10.023~~ & 9.966~ & $+0.057$ & 1.46  & 0.093  & 0.009 & 1.09 & 0.433  \\
$\Upsilon(2 ^3P_0)$ & 10.232~~ &10.268 ~& $-0.036$ & 2.37  & $-0.66~\,$ & 0.15&0.693 & 0.691  \\
$\Upsilon(2 ^3P_1)$ & 10.255~~ &10.316$^\sharp$ & ~$-0.061$$^\sharp$ & 3.97 & $-3.56~\,$ 
& 1.50 &0.552$^\sharp$ & 1.20  \\
$\Upsilon(2 ^3P_2)$ & 10.268~~ &10.457$^\sharp$ & ~$-0.189$$^\sharp$ & 4.55 & $-5.03~\,$ 
& 2.53 & 0.537$^\sharp$ & 1.39  \\
$\Upsilon(3 ^3S_1)$ & 10.355~~ &10.327~ & $+0.028$ & 2.34 & $-0.583$& 0.163&0.698 & 0.684  \\
$\Upsilon(4 ^3S_1)$ & 10.580~~ &11.760$^\sharp$ & ~$-1.180$$^\sharp$ & 5.45 & $-6.47~\,$ & 4.38 
&0.527$^\sharp$ & 1.61\\
\hline
$B_c(1^1S_0)$ & $6.4 \pm 0.4$ & 6.324~ & $0.08 \pm 0.4$ & 0.668~& 0.187 & 0.022 & 1.64 & 0.329 \\
\hline
\end{tabular}
\end{center}
\vspace{-3mm}
\caption{\footnotesize
Comparisons of the theoretical predictions of perturbative QCD and the experimental data.
We used $n_l=4$ for $b\bar{b}$ systems and $n_l=3$ for $c\bar{c}$ and $b\bar{c}$ systems.
All dimensionful numbers are in GeV unit.}
\label{table:spectra}
\end{table}

Using these masses as input and Eq. (\ref{scalefix}), we can calculate the energy levels 
of other observed quarkonium states. These are given in Tab. \ref{table:spectra}. 
The series expansions for the charmonium $1^1S_0$ state and for the $1^3S_1$, $1^3P_j$, $2^3S_1$, 
$2^3P_0$ and $3^3S_1$ bottomonium states converge well. For these states the differences 
of the theoretical predictions and the experimental data are typically  $30$--$70$ MeV.
Convergence of the series expansions is poor for the $2P_1$, $2P_2$ and $4S$ bottomonium states 
as well as for the other charmonium states. We consider that the theoretical 
predictions for these levels are unreliable and marked the corresponding theoretical predictions 
with sharps ($\sharp$) in the case of the bottomonium. In the charmonium case 
unreliable predictions are not displayed. The differences $E_X^{exp}-E_X$ are rather large 
for the states with sharps. Notice that for these states the corresponding $\als(\mu_X)$ 
becomes larger than one, indicating a breakdown of the perturbative series.
We also computed the mass of the $B_c(^1S_0)$ state. The theoretical prediction is consistent with 
the experimental value, although the experimental error is large.
Generally, for states, which we consider reliably calculable in the above perturbative 
approach, the scale dependence decreases as we include more terms of the perturbative series.
For states, whose predictions we consider unreliable, the series becomes much more convergent 
if we would choose a scale different from (typically larger than) $\mu_X$.

\begin{figure}[t]
  \hspace*{\fill}
    \includegraphics[width=10cm]{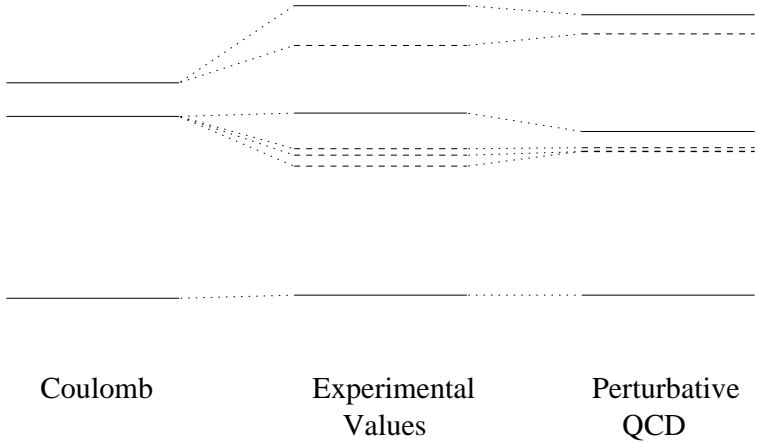}
  \hspace*{\fill}
\vspace{-7mm}
\\
\caption{\footnotesize
The bottomonium level structure as given by a pure Coulombic potential, by 
experiments and by the present analysis: the solid and dashed lines represent the $S$-states and $P$-states,
respectively. The input parameter of the perturbative QCD calculation is $\als^{(5)}(M_Z)=0.1181$ 
(see Tab. \ref{table:spectra}). We show only those levels that we can compute reliably. 
The Coulomb levels are calculated with $m_{\rm pole}=5.105$ GeV and $\als=0.5752$ 
such that they reproduce the $1S$ and $2S$ levels.}
\label{levelstr}
\end{figure}

The value of $\mu_X$ for the $1^3S_1$ state of $b\bar{b}$ is close to the central value (2.575 GeV) used 
in \cite{py}, while the value for the $1^3S_1$ state of $c\bar{c}$ is lower than the central value 
(1.57 GeV) used by the same authors. The value of $\mu_X$ used in Tab. \ref{table:spectra} 
for the $1^1S_0$ state of the $b\bar{c}$ is close to the central value (1.6 GeV) used in \cite{bcbv}. 
We also notice a remarkable agreement with their value of the $B_c$ mass.
This supports the idea that the scale of minimal sensitivity is, indeed, close to the 
characteristic scale of the system, which may be identified with the inverse of its size.

Fig. \ref{levelstr} compares the experimental values of the bottomonium spectrum with theoretical 
predictions after eliminating those states, which are not reliably calculable  
($2 ^3P_1$, $2 ^3P_2$ and $4 ^3S_1$ states). If we take an average of the $S$-wave and $P$-wave 
levels corresponding to each principal quantum number $n$, the theoretical predictions 
with $\als(M_Z)=0.1181$ reproduce the experimental values fairly well.
On the other hand, the predictions for the $S$--$P$ splittings and the fine splittings 
are smaller than the experimental values.

\section{Error Estimates} 
\label{s4}
Besides non-perturbative corrections, there are three different kinds of
uncertainties in our theoretical predictions for the quarkonium spectra, 
listed below and in Tab. \ref{error}. We examine them separately.
Also these examinations indicate that the theoretical predictions
for the $2^3P_1$, $2^3P_2$ and $4^3S_1$ bottomonium states are quite unstable, 
while the prediction for the $2^3P_0$ state is at the boundary. 
\begin{itemize}
\item[1)]{{\it Uncertainty of $\als^{(5)}(M_Z)$.}
We analyzed the quarkonium spectra varying the input parameter within the range given in \cite{pdg}:
$\als^{(5)}(M_Z)=0.1181 \pm 0.0020$. The level spacings become wider for larger $\als^{(5)}(M_Z)$,
which is consistent with our naive expectation. We find that the central value $\als^{(5)}(M_Z)=0.1181$ 
reproduces the whole level structure better than $\als^{(5)}(M_Z)=0.1201$ or 0.1161.}
\item[2)]{{\it Charm Mass Effects.}
We have also made an analysis of the bottomonium spectrum including finite charm mass effects.  
Since they deserve a detailed analysis of their own, including new calculations, we will report 
the full results in a separate paper \cite{bsv2}. Here we only summarize some
of the qualitative features of the 
effects and include them as a part of the uncertainties of our present analysis.
The corrections to the reliable predictions turn out to be positive and to become larger for higher states, 
ranging up to $\sim 200$ MeV.
Much of the effects, however,  may be reabsorbed by the uncertainties 
in $\als^{(5)}(M_Z)$ (as given above).\footnote{
We note that the sensitivities of the higher levels
to a variation of $\als^{(5)}(M_Z)$ increase by the charm mass
effects due to the decoupling of the charm quark.}
}
\item[3)]{{\it Uncertainties from Higher-Order Corrections.}
We take the maximum value of the following five estimates as an estimate of uncertainties 
from unknown higher-order corrections for each series expansion:
(i) 
The difference between the theoretical predictions computed
using $\als(\mu)$ as obtained by solving the renormalization-group equation 
perturbatively at 4 loops (Eqs. (3) and (11) of Ref. \cite{running})
and numerically at 4 loops (the data of Tab. \ref{table:spectra}).
(Note that the number of energy levels that can be determined in a reliable 
way is larger with the former definition of the running coupling constant. 
Also in that case, reliable predictions turn out to be close to the experimental values.)
(ii) 
The difference between the theoretical predictions computed using the 3-loop 
and the 4-loop (as in Tab. \ref{table:spectra}) running coupling constants,  
fixing $\als^{(5)}(M_Z) = 0.1181$.
(iii)
The difference between the theoretical estimates obtained by fixing $\mu_X$ on the minimum of  $|E^{(3)}|$ 
and the results of Tab. \ref{table:spectra}, obtained by fixing $\mu_X$ via the condition (\ref{scalefix}). 
(iv)
The contribution $\pm|E^{(3)}_X|$ from Tab. \ref{table:spectra}. 
(v)
For the $1S$ states we consider the ${\cal O}(\als^5 m)$ corrections 
calculated in the large-$\beta_0$ approximation in \cite{ks}.\\
For comparison, we list more conservative error estimates.
These are the variations of $\overline{m}_{b,c}$ and $E_X$
when we fix the scale as twice\footnote{
If we fix the scale as half of the minimal sensitivity scale,
$\mu = \mu_X/2$,
the predictions for the $n=2$ bottomonium
states appear to be meaningless, since the
scales are quite close to the infrared
singularity of the runnning coupling constant, and
the predictions for the $n \geq 3$ states do not exist, since the scales
lie below the infrared singularity.
}
of the minimal sensitivity scale:
$\mu = 2 \mu_X$, where $\mu_X$ is determined from Eq. (\ref{scalefix}).
}
\end{itemize}

\begin{table}[h]
\begin{center}
\begin{tabular}{c|l||r|r|r|r|r|r|r|r}
\hline
\multicolumn{2}{c||}{} & $\delta \als^{(5)}(M_Z) $ & 
\multicolumn{6}{c}{Estimates of higher-order corrections}\\
\cline{4-10}
\multicolumn{2}{c||}{} & $=\pm 0.0020$
& (i)~~ & (ii)~~ & (iii)\, & (iv) \, & (v) & $\pm$max & $\mu = 2 \mu_X$\\
\hline
\multicolumn{2}{c||}{$\delta \overline{m}_b$} & $^{-19}_{+18}$~ 
&$-2$~ & +1~ & 0~ & $\pm 7$~ & +5 &$\pm 7$~~ & +16~~\\
\multicolumn{2}{c||}{$\delta \overline{m}_c$} & $^{-16}_{+15}$~ 
&$-5$~ & +4~ &+3~ & $\pm 21$~ & +20 &$\pm 21$~~ & +37~~\\
\hline
& $\Upsilon(1 ^3P_0)$ & $^{+54}_{-48}$~ & +15~ & $-14$~ & 0~ & $\pm 3$~ & & $\pm 15$~ & $-53$~~\\
& $\Upsilon(1 ^3P_1)$ & $^{+63}_{-42}$~ & +22~ & $-8$~ & +7~ & $\pm 2$~ & & $\pm 22$~ & $-48$~~\\
& $\Upsilon(1 ^3P_2)$ & $^{+57}_{-50}$~ & +16~ & $-16$~ & 0~ & $\pm 3$~ & & $\pm 16$~ & $-54$~~\\
& $\Upsilon(2 ^3S_1)$~ & $^{+65}_{-58}$~ & +18~ & $-19$~ & $-1$~ & $\pm 9$~ & & $\pm 19$~ & $-73$~ \\
$\delta E_X$ & $\Upsilon(2 ^3P_0)$ & $^{+117}_{+755^\sharp}$ &+33~ & $-57$~ & $-21$~ & $\pm 150$~ & & $\pm 150$~ & $-120$~~\\
& $\Upsilon(2 ^3P_1)$ & $^{~~\, +83^\sharp}_{+1189^\sharp}$  & $-4$~ 
& $+1637^\sharp$ & $-66^\sharp$ & $\pm 1500^\sharp$~& & $\pm 1637^\sharp$ 
& $-97$~~\\
& $\Upsilon(2 ^3P_2)$ & $^{~~\, -44^\sharp}_{+1528^\sharp}$ & $-136$~ 
& $+2136^\sharp$ &$-206^\sharp$ & $\pm 2530^\sharp$~& &  $\pm 2530^\sharp$ 
& $-229$~~\\
& $\Upsilon(3 ^3S_1)$ & $^{+130}_{~ -93}$~ & +37~ & $-63$~ &$-36$~ 
& $\pm 163$~ & & $\pm 163$~ & $-161$~~\\
& $\Upsilon(4 ^3S_1)$ & $^{~\, -381^\sharp}_{+1883^\sharp}$ & $-639^\sharp$ 
& $+2936^\sharp$ &$-1425^\sharp$ & $\pm 4380^\sharp$& & $\pm 4380^\sharp$ 
& $-1361$~~\\
& $B_c(1 ^1S_0)$ & $^{+4}_{-5}$~ & $-4$~ & 0~ & $-1$~ & $\pm 22$~ & & $\pm 22$~& $+1$~~\\
\hline
\end{tabular}
\end{center}
\caption{\footnotesize
Variations of the theoretical predictions of Tab. \ref{table:spectra} 
when the uncertainties 1) and 3)
discussed in Sec. \ref{s4} are separately
taken into account. All dimensionful numbers are in MeV unit.
Those values corresponding to the unreliable theoretical predictions are marked with sharps. 
The input parameter is taken as $\als(M_Z)=0.1181$ except in the first column.
The column ``$\pm$max'' lists $\pm$max$\{|$(i)$|$, $|$(ii)$|$, $|$(iii)$|$, $|$(iv)$|$, $|$(v)$|\}$.
In the last column we write conservative estimates with the scale
choice $\mu = 2\mu_X$.}
\label{error}
\end{table}

\begin{figure}[t]
\begin{center}
\vspace{0mm}
\begin{minipage}{8.0cm}\centering
    \includegraphics[width=8cm]{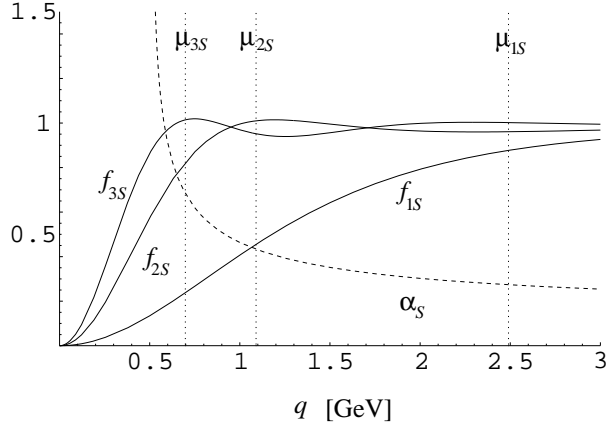}
\vspace{-3mm} \\
  \end{minipage}
\vspace{-7mm}
\end{center}
\caption{\footnotesize
The support functions $f_X(q)$ vs.\ $q$ for $X = 1S$, $2S$ and $3S$ (solid lines).
$f_X(q)$ is calculated using $m_{\rm pole}=5$~GeV and a different $\als(\mu_X)$, taken from 
Tab. \ref{table:spectra}, for each $X$. Vertical lines represent the corresponding scales $\mu_X$ taken
from the same table. Also $\als^{(4)}(q)$ is shown by a dashed line.
\label{fx2}}
\end{figure}

\section{Interpretation}
\label{s5}
\clfn
The most non-trivial feature of the present theoretical predictions for the bottomonium spectrum is that the level spacings between consecutive $n$'s are almost constant,
whereas in the Coulomb spectrum the level spacings decrease as $1/n^2$.\footnote{
If we consider, for instance, the ratio $(E_{3S}-E_{2S})/(E_{2S}-E_{1S})$, then we obtain 
experimentally $0.59$, from the data of Tab. \ref{table:spectra}, $0.71$, while 
from a pure Coulomb interaction $0.19$.} 
Conventionally, this same difference between the Coulomb spectrum
and the observed quarkonium spectra motivated people to construct
various potential models. It is, therefore, imperative to elucidate how
the above perturbative QCD calculation is able to reproduce such a level structure.
We will focus on two points:  (1) the leading renormalon cancellation, which implies that 
infrared physics decouples; 
this is essential to obtain convergent series expansions; (2) the meaning of the scale $\mu_X$ 
chosen by the scale-fixing prescription (\ref{scalefix}).

Let us approximate each term on the right-hand-side of Eq. (\ref{spectrum}) (in the equal-mass case) as
\bea
&&
2m_{\rm pole} \simeq 2 \overline{m} +
{\hbox to 18pt{ \hbox to -5pt{$\displaystyle \int$} 
\raise-19pt\hbox{$\scriptstyle |\vec{q}|< \overline{m}$} }}
\frac{d^3\vec{q}}{(2\pi)^3} \, | V_{\rm QCD}(q) |
= 2 \overline{m} + \frac{2 C_F}{\pi} \int_0^{\overline{m}} dq \, \alpha_V (q) ,
\label{poleapprox} \\
&& E_{{\rm bin},X} \simeq \bra{X} \frac{\vec{p}^{\, 2}}{m_{\rm pole}}+V_{\rm QCD} \ket{X}.
\label{binapprox}
\eea 
Here, $V_{\rm QCD}(q) = - C_F 4\pi \alpha_V(q)/q^2$ is the QCD static potential in momentum space;
$\ket{X}$ denotes the Coulomb wave function (the zeroth-order approximation) and $C_F=4/3$.
The first approximation follows from the fact that the dominant contribution to
the pole-$\overline{\rm MS}$ mass relation can be read from the infrared region, 
loop momenta $q \ll \overline{m}$, of the QCD static potential \cite{renormalon2}. 
The second approximation is more obvious. Let us note that also the right-hand-side of Eq. (\ref{binapprox})
incorporates and is dominated by the leading renormalon contribution to the static potential.
From Eqs. (\ref{poleapprox}) and (\ref{binapprox}) we obtain 
\begin{equation}
E_X \simeq 2 \overline{m} + \frac{2C_F}{\pi} \int_0^\infty dq \, \alpha_V (q) \, f_X(q) 
+ \bra{X} \frac{\vec{p}^{\,2}}{m_{\rm pole}} \ket{X} 
\simeq 2 \overline{m} + \frac{2C_F}{\pi} \int_0^\infty dq \, \alpha_V (q) \, f_X(q).  
\label{eth3}
\end{equation}
The last approximation follows from the fact that the kinetic energy contribution to the bottomonium 
levels turns out to be numerically small\footnote{It is about 
17\% of $E_X^{(1)}$ for the $X=1S$ state, 6\% of $E_X^{(1)}$ for the
$X=2S$ state, 4\% of $E_X^{(1)}$ for the $X=3S$ state. 
Moreover, these contributions tend to cancel each other in the 
level spacings.} 
(notice that this does not contradict the virial theorem, 
since the static potential we are considering here incorporates the running of $\als$ and therefore 
is not simply of the form $1/r$). The support function $f_X$ is 
\bea
f_X(q) = \theta (\overline{m}-q) - \int_0^\infty dr \, r^2 | R_X(r) |^2 \, \frac{\sin (qr)}{qr},
\eea
where $R_X(r)$ is the radial part of the Coulomb wave-function of $X$. 
$f_X(q)$ is almost unity in the region $1/a_X \simlt q < \overline{m}$, 
where $a_X$ is the inverse of the dumping scale of $f_X$ and may be
interpreted as the size of the bound-state $X$.\footnote{
According to Eq. (\ref{eth3}), $1/a_X$ acts as an infrared cut off in the
computation of the energy level. We may compare it with a qualitative 
picture where the gluons, whose wavelengths are much larger than the size 
of the color-singlet bound-state, cannot couple to it.}
For the $1S$ state $f_X(q)$ dumps slowly 
as $q$ decreases. For other states $f_X(q)$ dumps rapidly from scales which are somewhat smaller 
than the naive expectations $(C_F \als m_{\rm pole})/n_X$.
Eq. (\ref{eth3}) tells that  {\it the major contribution to the energy levels comes from the region 
$1/a_X$ $ \simlt$ $ q$ $\simlt$ $\overline{m}$ of the self-energy corrections of quark and antiquark}
(apart from the constant contribution $2 \overline{m}$).
In Fig. \ref{fx2} we show $f_X$ for different states calculated with $m_{\rm pole}=5$ GeV 
and $\als(\mu_X)$ taken from Tab. \ref{table:spectra}. The corresponding values 
of $\mu_X$ are also displayed. For those states which we consider the predictions reliable,
$\mu_X$ is located within the range where $f_X(q) \simeq 1$ (close to the infrared edge);
for those states with unreliable predictions, $\mu_X$ is located out of this range but far in the 
infrared region. In Fig. \ref{fx2} also $\als(q)$ is shown. We see that as $n_X$ 
increases from 1 to 3, the coupling $\als(q)$, close to the dumping scale of $f_X(q)$, grows rapidly.
According to Eq. (\ref{eth3}), this is the very reason for the widening
of the level spacings in excited states in comparison to the Coulomb spectrum.

Summarizing the above discussion we may draw the following 
qualitative picture of the structure of the bottomonium spectrum:
\begin{itemize}
\item
The energy levels of bottomonium are mainly determined by 
(i) the $\overline{\rm MS}$ masses of $b$ and $\bar{b}$, and 
(ii) contributions to the self-energies of $b$ and $\bar{b}$
from gluons with wavelengths $1/\overline{m} \simlt \lambda \simlt a_X$.
The latter contributions may be regarded as the difference between
the (state-dependent) constituent quark masses and the current quark masses.
\item
Level spacing between consecutive  $n$'s increases rapidly with $n$
as compared to the Coulomb spectra. This is because the self-energy contributions (ii)
grow rapidly as the physical size of the bound-state increases.
\end{itemize}

\section{Conclusions}
\label{s6}
For all the bottomonium states, where the predictions of perturbative QCD 
can be made reliably (i.e. $\als < 1$), our results are consistent with the experimental 
data within the estimated uncertainties of the theoretical predictions. The obtained value 
for $\overline{m}_b$ is in good agreement with the recent sum-rule calculations.
The theoretical uncertainties given in Tab. \ref{error} are numerically 
of the same size as $\Lambda_{\rm QCD}\times (a_X \Lambda_{\rm QCD})^2$,
i.e. of the effect of the ${\cal O}(\Lambda_{\rm QCD}^3)$
renormalons: if we approximate $1/a_X \simeq \mu_X$, take the values of 
Tab. \ref{table:spectra}, and $\Lambda_{\rm QCD} = 300 - 500$ MeV, we obtain 
for the $1S$ state a contribution of order $\pm(5 - 20)$ MeV, for the $n=2$ states a contribution 
of order $\pm(20 - 110)$ MeV and for the $3S$ state a contribution of order $\pm(50 - 250)$ MeV. 
Since the mass $\overline{m}_b$ has been fixed on the vector ground state and has {\it not} 
been adjusted for higher states, the data at our disposal suggest that:
1) the bulk of the bottomonium spectrum is accessible by perturbative QCD up to some 
of the $n=3$ states; 2) non-perturbative contributions do not need to be larger than $250$ MeV 
for the reliable $n=3$ states, than $100$ MeV for the $n=2$ states and than $20$ MeV for $\overline{m}_b$ 
and may be of the type associated with the ${\cal O}(\Lambda_{\rm QCD}^3)$ renormalon. 
These upper bounds to the non-perturbative corrections are conservative and their true sizes 
may be considerably smaller; note that for reliable predictions all of $|E_X^{exp} - E_X|$ 
in Tab. \ref{table:spectra} are smaller than 80 MeV. The existence of non-perturbative 
corrections of the above sizes may also explain the discrepancy observed 
in Tab. \ref{table:spectra} between the theoretical estimates and the experimental data of the fine 
structure and the $2S$-$1P$ splittings. However, before making any definitive statement, an analysis 
including higher-order corrections is necessary.

The reliability of the perturbative calculations and the agreement with the experimental 
data indicates that the assumptions made at the end of Sec. \ref{s2} are, indeed, satisfied by the 
$n=1$, $n=2$ and some of the $n=3$ bottomonium states.
Hence, non-perturbative contributions are encoded into non-local condensates, which 
may reduce to local ones for the ground states. These are expected to reabsorb 
the ${\cal O}(\Lambda_{\rm QCD}^3)$ renormalon \cite{long} and are of the order 
$\Lambda_{\rm QCD}\times (a_X \Lambda_{\rm QCD})^2$. 
The agreement noticed above, between this estimate and the uncertainties of the perturbative series 
given in Tab. \ref{error}, shows the consistency of our conclusions. 

For what concerns the $c\bar{c}$ system, with the present method we are not in the condition  
to make reliable predictions for states higher than the ground state, and, therefore, 
we cannot extrapolate from consistency arguments the size (and the nature) of the non-perturbative 
corrections. We have noticed, however,  that our estimate for $\overline{m}_c$ is in good agreement with 
recent sum-rule calculations. From the prediction of the $\eta_c$ mass we may guess non-perturbative 
contributions to be, in this case, of the order of $100$ MeV. Also this figure is consistent 
with the  ${\cal O}(\Lambda_{\rm QCD}^3)$ renormalon effect ($20 - 110$ MeV).

\section*{Acknowledgements}
Y.S.\ is grateful to fruitful discussions with K.~Hikasa and T.~Tanabashi. Y.S.\ was supported in part
by the Japan-German Cooperative Science Promotion Program.

\end{document}